\documentclass[12pt]{article}
\usepackage{amssymb}
\newcommand{\bq}{\begin{eqnarray}}
\newcommand{\eq}{\end{eqnarray}}
\newcommand{\ra}{\rightarrow}
\newcommand{\ov}{\overline}

\begin{document}

\begin{center}{\bf ON MIXED PHASES IN GAUGE THEORIES}\end{center}

\begin{center}{V.L. CHERNYAK}\end{center}

\begin{center}{Budker Institute for Nuclear Physics\\
 630090 Novosibirsk, Russia\\
E-mail: chernyak@inp.nsk.su}
\end{center}
\vspace{5mm}

\begin{center} 
Talk given at "Continuous  Advances in QCD-2002/Arkadyfest",\\ 
 honoring the 60-th birthday of ARKADY VAINSHTEIN; \\

\vspace{2mm}
17-23 May 2002, University of Minneapolis, Minnesota, USA
\end{center}

\vspace{6mm}

\begin{center}{\bf Abstract}\end{center}

In many gauge theories at different values of parameters entering
Lagrangian, the vacuum is dominated by coherent condensates of different 
mutually non-local fields (for instance, by condensates of electric or 
magnetic charges, or by various dyons). It is argued that the transition 
between these "dual to each other" phases proceeds through the intermediate 
"mixed phase", having qualitatively different features. The examples 
considered include: ordinary YM,\,\, $N=1\,\,$ SYM,\,\, $N=1\,\,$ SQCD,\, 
and broken $N=2\,$\, SYM and SQCD.     

\newpage
{\begin{flushright}{\bf To Arkady,\\my friend and teacher}\end{flushright}}
\vspace{5mm}

\begin{center}{\bf 1. {\boldmath $SU(N_c)$ - YM at $\theta\neq 0$} 
and dyons}\end{center}

The physics of this theory, and in particular the vacuum energy density
${\ov E}_{vac}(\theta,\,N_c)$, is supposed to be periodic in \, $\theta\ra 
\theta+2k\pi$. On the other hand, the standard large $N_c$-counting rules 
imply ($b_o=11/3$): 
\bq
{\ov E}_{vac}= - N_c^2\frac{b_o}{4}\Lambda^4\,F(\theta/N_c)\,\,,
\eq
with $F(z\ra 0)=1-c_1 z^2+c_2 z^4+\dots\,\,\,$
\footnote{
The numerical coefficient $c_1$ is positive, but it is a dynamical
quantity and can not be determined from general considerations alone.
}
$\,\,\,$ It was first pointed out by
E. Witten [1] (see also [2] for a similar behaviour in the "stringy-YM"
theory) that the $N_c$-dependence in Eq.(1) and periodicity in $\theta\ra 
\theta+2\pi$ imply together that the function $F(z)$ should be nonanalytic 
in its argument. So, for instance, instead of Eq.(1), the more explicit form
of dependence of ${\ov E}_{vac}$ on $\theta$ will rather look as:
\bq
{\ov E}_{vac}= - N_c^2\frac{b_o}{4}\Lambda^4\left \{min\,\sum_{k}\,f\left 
( \frac{\theta+2k\pi}{N_c}\right )\right \}\,,
\eq
with $f(z)$ being the "normal" analytic function.

The qualitative behaviour of the curve ${\ov E}_{vac}(\theta,\,N_c)$ 
looks as follows. First, it is symmetric in $\theta\ra -\, \theta$ and 
periodic in $\theta\ra \theta+2k\pi$. Further, it has  minimum at 
$\theta=0$ and begins to increase with increasing $\theta>0$, as it 
follows from general considerations of the Euclidean functional 
integral determining this theory. It reaches its maximal value at $\theta
=\pi$. The curve itself is continuous at this point, but there is a cusp 
so that ${\ov E}_{vac}$ begins to decrease in a symmetric way in the
interval $\pi <\theta <2\pi$, reaching the same minimal value at $\theta=
2\pi$. 

As for the qualitative behaviour of the topological charge density, 
${\ov P}_{vac}(\theta,\,N_c)$, it follows from the relation: ${\ov P}_{vac}
(\theta,\,N_c)\sim d{\ov E}_{vac}/d\theta$, and looks as follows.
First, it is antisymmetric in $\theta\ra -\, \theta$ and periodic in $\theta
\ra\theta+2k\pi$. So, it is zero at $\theta=0$ and increases with $\theta$
reaching its maximum value at $\theta=\pi$. There is a discontinuity at this
point, so that the curve jumps to the same but negative value as $\theta$
overshoots $\pi$, and increasing in a symmetric way reaches zero at
$\theta=2\pi$. 

The above described nonanalytic (cusped) behaviour of 
${\ov E}_{vac}(\theta,\,N_c)$ 
along the real $\theta$-axis agrees, in particular, with the asymptotic
behaviour of ${\ov E}_{vac}(\theta,\,N_c)$ at large imaginary values of 
$\theta, \,\,i\theta/N_c={\tilde \theta}/N_c \gg 1,\,$ obtained in [3]:
\bq
{\ov E}_{vac}(\theta,\,N_c)\sim -  N_c^2\,\Lambda^4\,\exp\,\left \{\frac{4}
{b_o}\,\frac{\tilde \theta}{N_c}\, \right \}\,.
\eq
It is seen from Eq.(3) that ${\ov E}_{vac}(\theta,\,N_c)$ is not naturally 
periodic at $\tilde \theta\ra \tilde\theta \pm2i\pi$. Rather, it implies 
that periodic ${\ov E}_{vac}(\theta,\,N_c)$ is analytic in the strip $-\pi< 
Re\,\theta <\pi$ in the complex $\theta$-plane, and is glued then 
periodically strip by strip.
\vspace{3mm}

The natural physical interpretation explaining the origin of the above
described cusped behaviour of ${\ov E}_{vac}(\theta)$ along the real
$\theta$-axis has been proposed in [3], and looks as follows.

Let us suppose the "standard" picture of the confinement mechanism to be 
valid, i.e. those of the dual superconductor. By this we imply here 
the dynamical mechanism with composite (naturally adjoint) Higgs field
which determines the formation of $U(1)^{N_c-1}$ 
from the original $SU(N_c)$,\,\, $SU(N_c)\ra U(1)^{N_c-1}$,\,\, 
and besides the $U_{i}(1)$-magnetically charged excitations (monopoles) 
condense. We will be interested to trace the qualitative behaviour of 
this vacuum state in its dependence of $\theta$. For this, it will be 
sufficient to consider the "first" U(1)-charge only with its monopoles, the 
dual photon and corresponding $g^{\pm}$ gluons as if it were the SU(2)
theory, because $\theta$ is $SU(N_c)$-singlet and all other U(1) charges 
will naturally behave the same way under variation of $\theta$.

As has been shown by E. Witten [4], the pure monopole M=(magnetic charge\,
=\,1,\, electric charge\,=\,0) at $\theta=0$ turns into the dyon with charges 
$d_1^{\theta}=(1,\,\theta/2\pi)$ at $\theta\neq 0$. 
So, the coherent 
condensate of monopoles in the vacuum at $\theta=0$ turns into the condensate 
of $d_1^{\theta}$-dyons as $\theta$ starts to deviate from zero, and the 
vacuum energy density begins to increase for this reason.

It is a specific property of our system that there are two types of
condensates made of the dyons and antidyons with the charges: 
$\{(1,1/2);\,(-1,-1/2)
\}$ and $\{(1,-1/2);\,(-1,1/2)\}$, and having the same energy density. 
This can be seen, for instance, as follows. Let us start from the pure
monopole condensate at $\theta=0$ and let us move anticlockwise along the 
path: $\theta=0\rightarrow \theta=\pi$. The vacuum state will consist of
$ (1,\frac{1}{2})$ - dyons and $ (-1,-\frac{1}{2})$ -
antidyons. Let us move now clockwise along the path: $\theta=0\rightarrow 
\theta=-\pi$. The vacuum state will consist now of
$ (1,-\frac{1}{2})$ - dyons and $ (-1,\frac{1}{2})$ -
antidyons. Because the vacuum energy density is even under $\theta\ra \,
-\theta$, these two vacuum states are degenerate.
\footnote{
The existence of two vacuum states at $\theta=\pi$ does not follow from the 
symmetry considerations alone, like ${\ov E}_{vac}(\theta)={\ov E}_{vac}(-
\theta)$ and ${\ov E}_{vac}(\theta)={\ov E}_{vac}(\theta + 2\pi k)$. 
It is sufficient to give a counterexample. So, let us consider
the SU(2) Yang-Mills together with the Higgs doublet with large 
vacuum condensate. In this case the $\theta$\,-dependence of the vacuum 
energy density is due to a rare quasiclassical gas of instantons, and is
$\sim \cos(\theta)$. All the above symmetry properties are fulfilled, but
there is only one vacuum state at $\theta=\pi$. (See also the end of this
section.)
}

Besides, these two states belong to the same world as they are reachable one 
from another through a barrier, because there are electrically charged gluons, 
$g^{\pm}=(0,\pm 1)$, which can recharge these $(1,\pm 1/2)$ 
- dyons into each other. In contrast, the two vacuum states, $|\theta 
\rangle $ and $|-\theta \rangle $ at $\theta \neq 0,\, \pi$ are 
unreachable one from another and belong to different worlds, as there is no
particles in the spectrum capable to recharge the $(1,\pm \theta/2\pi)$ -
dyons into each other. 

Thus, the vacuum state becomes twice degenerate at $\theta= \pi$, so that the 
"level crossing" (in the form of rechargement: $\{d_1=(1,1/2),\,{\bar d}_1=
(-1,-1/2)\}\ra \{d_2=(1,-1/2),\,{\bar d}_2=(-1,1/2)\}$)
can take place if this will lower the energy density at $\theta>\pi$. And 
indeed it lowers, and this leads to a casp in ${\ov E}_{vac}(\theta)$. At 
$\theta>\pi$ the vacuum is filled now with the coherent condensate of
new dyons with the charges: $d_2^{\theta}=(1,\,-1+\theta/2\pi),
\, {\bar d}_2^{\theta}=(-1,\,1-\theta/2\pi)$. 
As $\theta$ increases further, the electric charge of these $d_2^{\theta}$\,
-dyons decreases, and the vacuum energy density decreases with it. Finally, 
at $\theta=2\pi$ the $d_2^{\theta}$\,-dyons (which were the (1,\,-1)-dyons 
at $\theta=0$) become pure monopoles, and the vacuum state becomes 
exactly as it was at $\theta=0$, i.e. the same condensate of pure 
monopoles and antimonopoles.

We emphasize that, as it follows from the above picture, it is wrong to
imagine the vacuum state at $\theta=2\pi$ as, for instance, a condensate of 
dyons with the charges (1,\,-1), degenerate in energy with the pure 
monopole condensate at $\theta=0$.
\footnote{
In this respect, the widely used terminology naming the two singularity points
$u=\pm \Lambda^2$ on the $N=2\,\,\, SU(N_c=2)$ SYM moduli space as those 
where monopoles and respectively dyons become massless, is not quite adequate
(and may be dangerous for this reason, leading to wrong conclusions).
It is based on {\it quantum numbers} ${\vec n}=(n_m,\,n_e)$ of corresponding
fields, and these quantum numbers are always the same independently of the
point of the moduli space we are staying in, and are not direct physical 
observables. In contrast, the standard physical terminology is based on 
{\it charges} ${\vec g}=(g_m,\,g_e)$ which are the direct physical observables
because, by definition, the Coulomb interaction of two particles is 
proportional to product of their charges, not quantum numbers. In distinction
from {\it quantum numbers}, the values of {\it charges} depend on 
the point of the moduli space, due to Witten's effect.

To illustrate, let us start from the vacuum $u=\Lambda^2$ where, by
definition, the massless particles are pure monopoles and
let us move, for instance, along a circle to the point $u=-\Lambda^2$. On
the way, the former massless monopole increases its mass because it becomes 
the $d_1^{\theta}=(1,\,\theta(u)/2\pi)$ - dyon (here $\theta(u)/2\pi=Re 
\,\tau(u))$. At the same time, the former massive $d_2^o=(1,\,-1)$ - dyon 
diminishes its mass as it becomes the $d_2^{\theta}=(1,\,-1+\theta(u)/2\pi)$ 
- dyon. When we reach the point $u=
-\Lambda^2,$ i.e. $\theta(u)=2\pi$, the former dyon becomes massless just 
because it becomes the pure monopole here. So, an observer living in the
world with $u=-\Lambda^2$ will also see the massless monopoles (not dyons, 
and this is distinguishable by their Coulomb interactions between themselves
and with other dyons), exactly as those living in the world with $u=\Lambda^2$.
}

Physically, the above rechargement process will appear as a typical first
order phase transition. After $\theta$ overshoots $\pi$, in a space with the
coherent condensate of $d_1=(1,1/2)$ - dyons and ${\bar d}_1=(-1,-1/2)$ 
- antidyons, the bubbles will appear with the coherent condensate of $d_2=(1
,-1/2)$ - dyons and ${\bar d}_2=(-1,1/2)$- antidyons deep inside each bubble, 
and with a transition region surface (domain wall) through which the averaged
densities of two type dyons interpolate smoothly. These bubbles expand then 
over all the space through the rechargement process $d_1+{\bar d}_1\ra d_2+{
\bar d}_2$ occuring on a surface of each bubble. This rechargement can be 
thought as going through a copious "production" of charged gluon pairs $g^{
+}g^{-}$, so that the underlying processes are: $[d_1=(1,1/2)]+[g^-=(0,-1
)]\ra [d_2=(1,-1/2)]$ and $[{\bar d}_1=(-1,-1/2)]+[g^+=(0,1)]\ra [{\bar d}
_2=(-1,1/2)]$.
\\

Some analogy with the simplest Schwinger model may be useful at 
this point, in connection with the above described rechargement process.
Let us consider first the pure $QED_2$ without finite mass charged 
particles, and let us put two infinitely heavy "quarks" with the charges
$\pm \theta/2\pi$ (in units of some $e_o$) at the edges of our space. It is 
well known [5] that this is equivalent to introducing the $\theta$\,-angle
into the $QED_2$ Lagrangian. As a result, there is the empty vacuum at
$\theta=0$, and the long range Coulomb "string" at $\theta \neq 0$. The
vacuum energy density behaves as: ${\ov E}_{vac}(\theta)=C_o e_o^2\, 
\theta^2,\, C_o=const,$ at any $0\leq \theta < \infty$.

Let us add now some finite mass, $m\gg e_o$, and of unit charge $e_o$
field $\phi$ to the Lagrangian. When there are no external charges, this
massive charged field can be integrated out, resulting in a small charge 
renormalization. But when the above quarks are introduced, the behaviour 
of ${\ov E}_{vac}(\theta)$ becomes nontrivial. The charge of the external 
quark tends to 1/2 as $\theta$ approaches $\pi$. As $\theta$ overshoots 
$\pi$ it becomes preferable to produce a 
pair of $\phi$\,- particles, $\phi^{+}\,\phi^{-}$, from the
vacuum. They separate so that to recharge the external quarks: 
$(\pm 1/2) \ra  (\mp 1/2)$ (without changing the volume energy), and the 
external charges become equal $(\theta/2\pi-1)$ and $(-\theta/2\pi+1)$ at 
$\theta>\pi$. As a result of this rechargement, there appears a 
cusp in ${\ov E}_{vac}(\theta)$ and it begins to decrease at $\theta> \pi$, 
so that the former "empty" vacuum is reached at $\theta=2\pi$. Therefore,
the behaviour of ${\ov E}_{vac}(\theta)$ will be: ${\ov E}_{vac}(\theta)
=C_o e_o^2\cdot \{\,\min\,{_k}\, (\theta+2\pi k)^2\,\}$, so that ${\ov E}_
{vac}(\theta)=C_o e_o^2\, \theta^2$ at $0\leq \theta \leq \pi$, and ${\ov E}
_{vac}(\theta)=C_o e_o^2\, (2\pi-\theta)^2)$ at $\pi\leq \theta \leq 2\pi$.\\

Let us return however to our dyons. The above described picture predicts 
also a definite qualitative behaviour of the topological 
charge density, ${\ov P}(\theta).$  
At $0<\theta<\pi$, i.e. in the condensate of the $d_1^{\theta}=(1,\theta/
2\pi)$ - dyons and ${\bar d}_1^{\theta}=(-1,-\theta/2\pi)$ - antidyons, the
product of signs of the magnetic and electric charges is positive for both 
$d_1^{\theta}$ - dyons and ${\bar d}_1^{\theta}$ - antidyons. Thus, these
charges give rise to the correlated field strengths: ${\vec E}\,
\Vert \,\theta {\vec H},\,\,\, {\vec E \cdot \vec H} > 0$, and both
species contribute a positive amount to the mean 
value of the topological charge density, so that ${\ov P}_1(\theta)>0$ and 
grows monotonically with $\theta$ in this interval following increasing 
electric charge $\sim \theta/2\pi$ of the dyon. 

On the other side, at $\pi<\theta<2\pi$, i.e.
in the condensate of the $d_2^{\theta}=(1,-1+\theta/2\pi)$ - dyons and 
${\bar d}_2^{\theta}=(-1,1-\theta/2\pi)$ - antidyons, the
product of signs of the magnetic and electric charges is negative for both 
$d_2^{\theta}$ - dyons and ${\bar d}_2^{\theta}$ - antidyons. Thus,
both species contribute a negative amount to ${\ov P}_2(\theta)$, such that: 
${\ov P}_2 (\theta)=-{\ov P}_1 (2\pi-\theta)$, and ${\ov P}(\theta)$ jumps 
reversing its sign at $\theta=\pi$ due to  rechargement. 

On the whole, it is seen that the cusped behaviour of ${\ov E}_{vac}(\theta)$ 
and discontinuous behaviour of ${\ov P}(\theta)$ appear naturally in this 
picture of the confinement mechanism in $SU(N_c)$ - YM theory, and are 
exactly the same that are expected from simplest general considerations and
were described in the beginning of this section.\\

Clearly, at $0\leq\theta<\pi$ the condensate made of only the $d_1^{\theta}
=(1,\,\theta/2\pi)$ - dyons (recalling also for a possible charged gluon 
pair production) can screen the same type $d_k^{\theta}=
[\,const\,(1,\,\theta/2\pi)+(0,\,k)\,]$ - test dyon only $(k=0,\pm 1,
\pm 2, \dots\,$;\, and the same for the $d_2^{\theta}$ - dyons at $\pi< 
\theta \leq 2\pi\,)$. So, the heavy quark-antiquark pair will be confined at 
$\theta\neq \pi$.

New nontrivial phenomena arise at $\theta=\pi$. Because there are two
degenerate states, i.e. the condensates of $(1,\pm \frac{1}{2})$ - dyons 
(and antidyons), a "mixed state" configuration becomes possible with, for 
instance, each condensate filling a half of space only, and with the domain 
wall interpolating between them. The simplest reasonings about the energy 
scales involved in this domain wall are as follows. The masses of 
relevant gluons $g^{\pm}=(0,\pm 1)$ and both dyons $(1,\pm 1/2)$ 
coexisting together in the core of the domain wall
are naturally $\simeq \Lambda_{YM}$, and so of the same size will 
be increase in energy density. Besides, there are $(N_c-1)$ independent 
$U_i(1)$ charges. 
%$d_{1,i}=(1,\,1/2)_i$\,\, and\,\, $d_{2,i}=(1,\,-1/2)_i$\,\, dyons,
On the whole, therefore, the domain wall tension will be $T\sim N_c\Lambda
^3_{YM}$, while its typical width will be $\sim 1/\Lambda_{YM}$.\\

Physically, the above domain wall represents "a smeared rechargement", i.e.  
smeared over space interpolation of electrically charged degrees of freedom 
between their corresponding vacuum values, resulting
in a smooth variation of the averaged densities of both type dyons 
$(1,\pm 1/2)$ through the domain wall. Surprisingly, there is no confinement 
inside the core of such domain wall. 

The reason is as follows. Let us take the domain wall interpolating along 
the z-axis, so that at $z\ra -\infty$ there is the main coherent density 
of $d_1=(1,\,1/2)$ - dyons, and at $z\ra \infty$, - that of $d_2=(1,\,-1/2)
$ - dyons. As we move from the far left to the right, the density of 
$d_1$-dyons decreases and there is also a smaller but increasing 
incoherent density of $d_2$-dyons. This small 
amount of $d_2$-dyons is "harmless", in the sense that its presence does 
not result in the screening of the corresponding charge. The reason is clear: 
the large coherent density of $d_1$-dyons confines the $d_2$-dyons 
so that they can not move freely and appear only in the form of
rare and tightly connected neutral pairs ${\bar d}_2d_2$, with different 
pairs fluctuating independently of each other. As we are going further to the
right, the density of these neutral pairs grows and their typical size 
increases (although they are still confined), because the main 
density of $d_1$-dyons decreases.  Finally, at some distance from the 
centre of the wall the percolation takes place, i.e. the $d_2$-dyons form 
a continuous coherent network and become released, so that the individual 
$d_2$-dyon can travel freely to arbitrary large distances (in the transverse 
xy-plane) but only within its network. And in this 
percolated region the coherent network of $d_1$-dyons still survives, so that 
{\it these two coherent networks coexist in space and form the new "mixed 
phase" with qualitatively different properties}. 

This is a general feature, and each time when there will coexist 
coherent condensates of two mutually non-local fields, 
they will try to confine each other, and will resemble the 
above described case.

The above mixed phase shares some features in common with the mixed state of 
the type-II superconductor in the external magnetic field. The crucial 
difference is that the magnetic flux is sourceless inside the superconductor, 
while in the above described mixed phase there are real dual to each other 
charges, each type living within its network.\\ 

As we move further to the right, the density of $d_2$-dyons 
continue to increase while those of $d_1$ continue to decrease. Finally, 
at the symmetric distance to the right of the wall centre the "inverse
percolation" takes place, i.e. the coherent network of $d_1$-dyons decays
into separate independently fluctuating neutral droplets whose average 
density (and size) continue to decrease with increasing $z$. Clearly, the 
picture on the right side repeats in a symmetric way those on the left one, 
with the $d_1$ and $d_2$ dyons interchanging their roles.
\footnote{
Evidently, if we replace the above domain wall with fixed $\theta=\pi$ by the
domain wall of the light axion field $a(z),\, m_a\ll \Lambda_{YM}$, 
interpolating between $a=0$ at $z\ra -\infty$ and $a=2\pi$ at $z\ra \infty$ 
with $a(z=0)=\pi$, all the properties will remain the same in the core, i.e.
at $|z|\lesssim (several)\, \Lambda_{YM}^{-1}$. The main difference will 
be that the condensates of $d_1=(1,\,1/2)$ and $d_2=(1,\,-1/2)$ - dyons will 
turn into condensates of pure monopoles at corresponding sides of large 
distances $|z|\gg 1/m_a$ (see also the next section). 
}

Let us consider now the heavy test quark put inside the core of the domain 
wall, i.e. inside the mixed phase region. Clearly, this region has the 
properties of the "double Higgs phase". Indeed, because the (two-dimensional)
charges of two dyons, $(1,\,1/2)$ and $(1,\,-1/2)$, are linearly 
independent, polarizing itself appropriately this system of charges will 
screen any external charge put inside, and the quark one in particular.  

Finally, if the test quark is put far from the core of the wall, the string
will originate from this point making its way toward the wall, and will be 
screened inside the mixed phase (i.e. the double Higgs) region. It should be 
emphasized however that, as it is clear from the above explanations, if this
test quark is moved further inside the core of the another wall, then its 
flux will be screened and nothing will support this string. So, 
the electric string can not be stretched between two such domain walls. 
%But if in the $\theta\neq \pi$ - vacuum 
%the finite size ball surrounding a quark and consisting of any 
%mixture of dyons and antidyons of any possible kind is excited, it will be 
%unable to screen the quark charge as there is no border at infinity where the
%residual polarization charge will be pushed out.
%\footnote{
%We have to make a reservation about the above described picture. There are 
%some reasons to expect that the point $\theta=\pi$ is very especial for $SU
%(N_c=2)$\,\,(\,unlike $SU(N_c\geq 3)$\,), because the
%vacuum energy density is likely to be exactly zero here. In this case, it
%is natural if the dyon condensate also approaches zero therein. The low
%energy theory 
%is expected then to have massless dyons etc., and to be rather a kind of a
%conformal field theory. Clearly, there will be no confinement in this phase.
%}
\\
  
Let us point out finally that the assumption about the confinement property
of the $SU(N_c)$ YM theory is not a pure guess, as
the above discussed nonanalytic (i.e. cusped) behaviour of the vacuum
energy density, ${\ov E}_{vac}(\theta,N_c)$, is a clear evidence for a phase
transition at some finite temperature. Indeed, at high temperatures the 
$\theta$ - dependence of the free energy density in the gluon plasma is 
under control and is\,: $\,\, \sim T^4(\Lambda/T)^{N_c b_o}\cos \theta$~,
\, due to rare gas of 
instantons. It is important for us here that it is perfectly analytic in 
$\theta$, and that the form of its $\theta$ dependence is T-independent, 
i.e. it remains to be  $\sim \cos \theta$ when the temperature decreases. 
On the opposite side at $T=0$, i.e. in the confinement phase, the $\theta$ -
dependence is nonanalytic and, clearly, this nonanalyticity survives at
small temperatures as there are no massless particles in the spectrum. 
So, there should be a phase transition (confinement -
deconfinement) at some critical temperature, $T_c\simeq \Lambda$, where the 
$\theta$ - dependence changes qualitatively.\\

\begin{center}{\bf 2. \hspace{5mm} {\boldmath $N=1\,\,\,SU(N_c)$}\,\,\, 
SYM} \end{center} 

In this theory the residual non-anomalous discrete axial symmetry is broken
spontaneously, so that there are $N_c$ vacuum states differing by the
phase of the gluino condensate [6],\,[7]:
\bq
{\langle 0 |\,\lambda\lambda\,| 0 \rangle}_k \sim N_c\Lambda^3\,\exp \left
 \{ i\frac{2\pi k}{N_c}\right \}\,.
\eq
Besides, it is widely believed that this theory is confining, similarly to
the usual YM- theory. In what follows, we will suppose that the confinement
mechanism here is the same as in the previous section, i.e. those of the 
dual superconductor.  Our purpose in this section will be to describe 
qualitatively the physical properties of domain walls interpolating between 
the above vacua and, in particular, their ability to screen the 
quark charge [8].

For this, let us consider the effective theory obtained by integrating out
all degrees of freedom except for the composite chiral field $S=(
W^2_{\alpha}/32\pi^2N_c),\,\,S=(\lambda\lambda,\cdots)/32\pi^2 N_c=(\rho
\exp\{i\phi\},\dots)$ (i.e. the integration proceeds with the constraint 
that the field $S$ is fixed, [8]). 
In this set up, the form of the superpotential in the 
Lagrangian for the field $S$ can then be simply obtained and
coincides with those of Veneziano-Yankielowicz [6]: $W\sim S\ln (S^N/\Lambda^
{3 N})$, resulting in the gluino condensation, see Eq.(4).

Because the field $N_c\phi$ in SYM is the exact analog of $\theta$ in the
ordinary YM, the physical interpretation and  qualitative behaviour of ${\ov 
E}_{vac}(\theta)$ in the YM- theory described in the previous section can 
be transfered now to SYM, with only some evident changes:\\ 
a) ${\ov E}_{vac}(\theta)\ra U(N_c\phi)$, and it is not the vacuum energy 
density now but rather the potential of the field $\phi$\,; \\
b) if we start with the condensate of pure monopoles at $\phi=0$, the
rechargement $d_1^{\phi}=(1,\,N_c\phi/2\pi)\ra d_2^{\phi}=(1,\,-1+N_c
\phi/2\pi)$ and the cusp in $U(N_c\phi)$ will occur now at $\phi=\pi/N_c$, 
so that at $\phi=2\pi/N_c$ we will arrive at the next vacuum with the same 
pure monopole condensate but with shifted phase of the gluino condensate.  

Let us consider now the domain wall interpolating along $z$-axis between two
nearest vacua with $k=0$ and $k=1$, so that $\phi(z)\ra 0$ at $z\ra -\infty$, 
and $\phi(z)\ra 2\pi/N_c$ at $z\ra \infty$. There is a crucial difference 
between this case and those described just above where the field $\phi$ was 
considered as being space-time independent, i.e. $\phi(z)=const$. The 
matter is that the system can not behave now in a way described above (which 
allowed it to have a lowest energy $U(N_c\phi)$ at each given value of 
$\phi(z)=\phi=const$): i.e. to be the pure coherent condensate of $d_1^{\phi}
$ - dyons at $0\leq \phi<\pi/N_c$, the pure coherent condensate of $d_2^
{\phi}$ - dyons at $\pi/N_c < \phi\leq 2\pi/N_c$, and to recharge suddenly 
at $\phi=\pi/N_c$. 
The reason is that the fields corresponding to electrically charged degrees 
of freedom also become functions of $z$ at $q=\int dz\, [ d\phi_{dw}(z)/dz ]
\neq 0$. So, they can not change abruptly now at $z=0$ where $\phi_
{dw}(z)$ goes through $\pi/N_c$, because their kinetic energy will become 
infinitely large in this case. Thus, 
the transition will be smeared necessarily.

The qualitative properties of the domain wall under consideration here will 
be similar to those described in the previous section. The main difference 
is that $\theta$ was fixed at $\pi$ in sect.1, while $N_c\phi_{dw}(z)$ acts
like the axion field, i.e. it varies here 
between its limiting values, and the electric charges of dyons follow it.

So, at far left there will be a large coherent condensate of $d_1^{\phi}=
(1,\,N_c\phi/2\pi)$-dyons (pure monopoles at $z\ra -\infty$), and a small 
incoherent density of $d_2^{\phi}=(1,\,-1+N_c\phi/2\pi)$-dyons.
\footnote{
Other possible dyons play no role in the transition we consider, and we will
ignore them.
}
The $d_2^{\phi}$-dyons can not move freely in this region as they are 
confined, and appear as a rare and tightly connected neutral pairs ${\ov d
}_2^{\phi}d_2^{\phi}$ only. Therefore, their presence does not result in the 
screening of the corresponding charge. As we move to the right, the density
of $d_1^{\phi}$-dyons decreases while those of $d_2^{\phi}$ - increases.
These last move more and more freely, but are still confined.
Finally, their density reaches a critical value at $z=-z_o$, so that 
a percolation takes
place and the $d_2^{\phi}$-dyons form a continuous coherent network within 
which the individual $d_2^{\phi}$-dyons can move freely to any distance
(in the transverse xy-plane). At the same time, there still survives a 
coherent condensate of $d_1^{\phi}$-dyons, which still can freely move 
individually within their own network. 

At the symmetrical point $z=z_o$ to the right of the domain wall centre at
$z=0$, the "inverse percolation" takes place, so that the coherent
connected network of $d_1^{\phi}$-dyons decays 
into separate independently fluctuating neutral droplets, whose density 
(and size) decreases with further increasing z. At large z we arrive at the 
vacuum state with a large coherent condensate of monopoles 
(former $(1,\,-1)$-dyons at large negative z). 

Now, let us consider what happens when a heavy quark is put inside the core 
of the domain wall. The crucial point is that there is a mixture of all four 
dyon and antidyon species (of all $N_c-1$ types): $d_1^{\phi}=(1,\,N_c\phi/2
\pi),\, {\bar d}_{1}^{\phi}=(-1,\,-N_c\phi/2\pi),\,d_2^{\phi}=(1,\,-1+N_c
\phi/2\pi)\,$ and ${\bar d}_2^{\phi}=(-1,\,1-N_c\phi/2\pi)$ in this 
percolated region, with each dyon moving freely inside its coherent
network. So, this region has the properties of "the double Higgs phase", as
here both the $d_1^{\phi}$ and $d_2^{\phi}$-dyons are capable to screen 
corresponding charges. And because the charges of $d_1^{\phi}$ and $d_2^
{\phi}$-dyons are linearly independent, polarizing itself appropriately
this mixture of dyons will screen any test charge put inside, the heavy 
quark one in particular.

If the test quark is put at far left (or right) of the wall, the string will 
originate from this point making its way towards a wall, and will disappear 
inside the core of the wall, i.e. in the mixed phase region where the
string flux will be screened. And similarly to the previous section, the
string can not be stretched between two domain walls.
 
The above described explanation of the physical phenomena resulting in
quark string ending in the wall differs from both, those described by E. 
Witten in [9] and those proposed by I. Kogan, A. Kovner and M. Shifman 
in [10] (see also the footnote 3).\\

\begin{center}{\bf 3. \hspace{5mm} {\boldmath $N=1\,\,\,SU(2)\,,\,\,N_F=1$} 
\,\, SQCD} \end{center}

As previously, we will imply here that there is confinement of electric
charges in the $N=1$ pure SYM- theory (see previous sections). Then, there 
will be three phases in this $N=1$\,\,SQCD - theory, depending on the value 
of $m$ - the mass parameter of the quark [8]. 

At small $m \ll\Lambda$, there will be the usual electric Higgs phase, 
with the large quark condensate $\langle {\ov Q}Q\rangle \sim (\Lambda^5/m)
^{1/2},$ and light quark composite fields $({\ov Q}Q)^{1/2}$ with masses 
$\sim m$. The effective low energy Lagrangian for these fields is those of
Affleck-Dine-Seiberg [11].

The heavy magnetically charged excitations (monopoles) will be confined, 
and so the monopoles will appear as rare and tightly bound neutral pairs 
only, with different pairs fluctuating independently of each other. 
\footnote{
That there are monopoles in this theory at $m\ll \Lambda$ can be seen as 
follows. First, let us consider the effective Lagrangian obtained by
integrating out hard degrees of freedom with high energy scales 
$\mu \geq \mu_o,\,\mu_o\sim (\Lambda^5/m)^{1/4}$. 
These include the instanton contributions, as the typical instanton size is 
${\ov \rho}\sim \langle {Q^\dagger}Q\rangle ^{-1/2}\sim 1/\mu_o$.
The instanton will add the Affleck-Dine-Seiberg term 
$\Lambda^5/({\ov Q}Q)$ to the original superpotential $m{\ov Q}Q$. 
Now, the so obtained effective Lagrangian is the appropriate
one to look, in particular, for a possible string solution if the 
characteristic distances involved in the string formation are 
larger than $\ov \rho$.

This is the case at the classical level, and there will 
be the solution for the 
Abrikosov-Nielsen-Olesen like string with the magnetic flux. But because the 
quarks are in the fundamental representation, the gauge group is SU(2) which
is simply connected and there are no truly
uncontractable strings in this theory. This implies that the above classical
string will break up on account of quantum tunneling effects. Physically,
this break up will be realized through the production of a pair of
magnetically charged "particle" and "antiparticle", with their subsequent 
separation along the string axis to screen the external infinitely heavy 
monopoles at the string ends. So, these magnetically charged particles 
should be present in the excitation spectrum of this theory (even if they 
are not well formed). 
}

With increasing $m$ the quark condensate and the monopole mass decrease, 
while the density of monopole pairs 
and their typical size increase. At some value $m=c_1\Lambda$ 
the percolation of the monopole droplets takes place, so that in the 
interval\, $c_1\Lambda \leq m \leq c_2\Lambda\, $ there will be the mixed 
phase (or equivalently, "the double Higgs phase") with two infinite size 
connected coherent networks of monopoles and quarks, with their averaged 
densities being constant over the space and following only the value of $m$.

There will be screening rather than confinement (although the difference 
between these two becomes to a large extent elusory here) of 
any test charge in this interval of $m$. 

Finally, at $m=c_2\Lambda$ the quarks become too heavy and can not support 
their coherent condensate anymore, so that this last decays into
independently fluctuating neutral droplets whose density and typical size
decrease with increasing $m$. 

At $m\gg \Lambda$ we arrive at the $N=1\,\,
SYM$ - theory with $\Lambda_{YM}=m^{1/6}\Lambda^{5/6}$, and with heavy 
($m\gg \Lambda_{YM}$) quarks which are confined.

The chiral quark condensate $\langle {\ov Q}Q\rangle \sim \Lambda^{5/2}/m^
{1/2}\sim\Lambda_{YM}^3/m$ is small but nonzero even in this region, but 
this small value is unrelated here with the gluon masses and charge screening 
by quarks, and is a pure quantum loop effect of heavy quarks: $\langle {\ov Q}
Q\rangle \sim \langle \lambda\lambda \rangle /m$ (similarly to the heavy
quark condensate $\langle {\ov \Psi}\Psi \rangle \sim \langle G_{\mu\nu}^2
\rangle / m$ in the ordinary QCD).

\begin{center}{\bf 4. \hspace{5mm} Broken {\boldmath $N=2\,\,\,SU(2)$} 
SYM} \end{center}

Let us recall the famous solution of this theory by N. Seiberg and
E.~ Witten, with the low energy Lagrangian (at small $\mu\ll \Lambda$)\,[12]:
\bq
L=\int d^4\theta\, \{ M^{\dagger}e^{V_D} M+{\ov M}^{\dagger}e^{-V_D} {\ov M}
+Im\,(A_D A^{\dagger}/4\pi)\}
-\frac{i}{16\pi} \int d^2\theta\,\tau_D\,{W}_D^2+\nonumber
\eq
\bq
+\int d^2\theta\,\{\sqrt 2\,{\ov M} M A_D+\mu\, U(A_D)\}+\,h.c.
\eq

Here $M$ is the monopole field. Because it was not integrated yet, the terms
entering the Lagrangian in Eq.(5) ($\tau_D$, etc.) do not contain the monopole
loop contributions and have no singularity at $\mu\ra 0$ and $\langle U 
\rangle \ra \Lambda^2$. The field $V_D$ is those of the dual photon 
and $W_D$ is its field strength.  
Below, it will be convenient for us to consider $A$ and $U$ in Eq.(5) 
as functions of the field $A_D$ which is a pure quantum field, i.e. has zero 
vacuum expectation value. 
The vacuum state we are dealing with is at $\langle U \rangle =\Lambda^2$, 
with $\tau_D\sim 1$ and $Im\,(A_D A^{\dagger}/4\pi)\sim A_D A^{\dagger}_D$ 
in Eq.(5) at small $\mu$. 

How the effective Lagrangian for these fields can look if $\mu$ is large
in comparison with $\Lambda$ ? Because at $\mu\gg\Lambda$ the
degrees of freedom which have been integrated out were heavy (in particular,
the charged Higgs fields with their masses $\sim \mu$, and charged $W^{\pm}$
\, bosons with their masses $m_W\sim \Lambda_{YM}=\mu^{1/3}\Lambda^{2/3}$\,), 
\,the $N=2$ supersymmetry will be broken explicitly and $\mu$- dependence 
will penetrate the effective Lagrangian. At
the same time, it is not difficult to see that due to: a) holomorphicity;\,
b) $R$-charge conservation (with the $R$-charge of $\mu$ equal two);\, c)
the known limit at $\mu\ll\Lambda$, the additional $\mu$- dependence can not
appear in the $F$ - terms, and so will appear in the $D$ - terms only.

Besides, restricting ourselves to the terms with no more than two space-time 
derivatives, it will be sufficient for our purposes to write these D-terms 
for the monopole and $A_D$ fields as the standard kinetic terms multiplied by
the c-number Z-factors $Z_{M}(\mu/\Lambda)$ and $Z_{H}(\mu/\Lambda)$, 
originating from those degrees of freedom which have been integrated out. 
%\footnote{
%For instance, there will be contributions to $Z_M$ in powers of $\langle M^
%{\dagger}M\rangle / M_W^2$ which are small $\sim (\mu/\Lambda)$ at $\mu\ll
%\Lambda$, and become large $\sim (\mu/\Lambda)^{1/3}$ at $\mu \gg \Lambda$. 
%}
Let us denote by $L_{\mu}$ the so obtained Lagrangian.

Recalling that the original theory was $N=2$ SYM broken by the mass
term of the Higgs fields, we are ensured that at $\mu\gg\Lambda$ the Higgs
fields become heavy, with their masses $m_H \sim \mu$, and decouple. 
So, we end up with $N=1$ SYM with the scale parameter: $\Lambda_{YM}=
\mu^{1/3}\Lambda^{2/3}$, and this is the only scale of this theory.

On the other hand, one obtains from $L_{\mu}$ at $\mu\gg\Lambda$ that 
\footnote{
In the above described set up, there is no need for the function $U(A_D)$ 
of the $\mu\, U(A_D)$ term in Eq.(5) to be exactly the Seiberg-Witten 
function. For our purposes and for simplicity, it will be sufficient
to keep only three first terms (i.e. constant, linear and quadratic in
$A_D$) in the expansion of $U(A_D)$ in powers of $A_D$, to ensure that the
adjoint Higgs becomes heavy, $m_{H}\sim \mu$, and decouples.
}
$Z_H$ stays intact, $Z_H(\mu/\Lambda\gg 1)\sim 1$, in order to have $m_H
\sim \mu$, while the values of the dual photon and monopole masses look as
\bq
m_{\tilde \gamma}^2 \sim Z_{M}(\mu/\Lambda)\,\langle 0|
{M^\dagger}M|0\rangle \sim Z_{M}(\mu/\Lambda)\cdot \mu
\Lambda\,,\quad m_{M}^2\sim Z_{M}^{-2}(\mu/\Lambda )\cdot\Lambda^2\,.
\eq

Let us combine now Eq.(6) with the additional assumption: {\it there is no
massless particles in the spectrum of}\, $N$=1\,\, SYM. Then, this requires
\footnote{
If $(\mu/\Lambda)^{1/3}\cdot Z_{M}(\mu/\Lambda)\ra 0,$ the dual photon will 
be massless (on the scale $\Lambda_{YM}$, i.e. $m_{\tilde \gamma}/\Lambda_
{YM}\ra 0)$,\, while if $(\mu/\Lambda)^{1/3}\cdot Z_{M}(\mu/\Lambda)\ra 
\infty$ the monopole will be massless.
}
:
\bq
Z_{M}(\mu/\Lambda )\sim \left (\frac{\Lambda}{\mu}
\right )^{1/3}\quad at \quad \mu\gg \Lambda\,.
\eq

It follows now from Eqs.(6),\,(7):
\bq
m_{\tilde \gamma}\sim m_{M}\sim \Lambda_{YM}\quad at \quad \mu\gg\Lambda\,,
\eq
i.e. {\it both the dual photon and monopole survive in the spectrum of the}
$N=1$\,\, SYM. This is nontrivial in the sense that one of them or both could 
become heavy and decouple at $\mu\gg\Lambda$.

As for the value of the monopole condensate, it depends clearly on the 
normalization of the monopole field. In the presence of the monopole $Z_{M}$
factor, the old normalization $\langle 0|{\ov M}M|0\rangle \sim \mu\,\Lambda$ 
is not the natural one. The appropriate normalization is: $\langle 0|{
\ov N}N|0\rangle =\langle 0|Z^{1/2}_M{\ov M}\cdot Z^{1/2}_M M|0\rangle$, 
and it has the right scale: $\langle 0|{\ov N}N|0\rangle \sim 
\Lambda^2_{YM}$.
\footnote{
It is not difficult to trace the role of higher order terms $\sim A_D^2(A_D/
\Lambda)^{k\geq 1}$ in the expansion of $U(A_D)$ in $L_\mu$ 
in powers of $(A_D/\Lambda)$. After integrating out the heavy 
field $A_D$, these will give additional terms in the superpotential in 
powers of the pure quantum field $({\ov M}M-\mu\Lambda)/\mu\Lambda$.
Rescaling the monopole field to $N=Z_M^{1/2}\,M$, to have the canonical 
kinetic term for the field $N$, these additional terms in the superpotential 
will come in powers of the pure quantum field $({\ov N}N-\Lambda^2_{YM})/
\Lambda^2_{YM}$, i.e. depending only on the scale $\Lambda_{YM}$, as it 
should be in the $N=1$ SYM theory.
}

On the whole, the above described results were obtained implying that
there is no phase transition in the broken $N=2$ SYM theory when going from 
small $\mu\ll\Lambda$ to large $\mu\gg\Lambda$ (in a sense, at least, that 
monopoles continue to form the coherent condensate which gives the mass to
the dual photon; on the other hand, restructuring of the spectrum definitely
occurs at $\mu\sim \Lambda$). They show a selfconsistency of 
this assumption and give a strong support to the
widely accepted expectation that the $N=1$ SYM theory is confining, with 
the confining mechanism those of the dual superconductor. In other words,
when going from $\mu\ll\Lambda$ to $\mu\gg\Lambda$ in the broken $N=2\,\,
\, SYM$ - theory,  the "external" adoint Higgs of this theory decouples 
at $\mu\sim \Lambda$ and its role is taken by the dynamically formed and
condensing  "internal" composite adjoint Higgs of the $N=1\,\, SYM$ - theory,
while monopoles continue to form the coherent condensate and keep the dual 
photon massive.\\

\begin{center} {\bf 5. \hspace{5mm} Broken}  \boldmath { $N=2\,\,\,SU(2),\,
\,N_F=1\,\,\, SQCD$ } \end{center}

The solution of the unbroken $N=2$ theory has been given by N. Seiberg
and E. Witten [13]. The original superpotential of the broken $N=2\ra N=1$ 
theory has the form (the kinetic terms are canonical):
\bq
W=m\,{\ov Q}Q+h\,\sqrt{2}\,\,{\ov Q}\,\frac{\tau^a}{2}\phi^a\,Q\,+
\mu\,\phi^2\,,
\eq
where the quark fields $Q$ and $\ov Q$ are in the $\bf 2$ and ${\ov {\bf 2}}
$ representations of the colour group $SU(2)$, and $\phi$ is the adjoint 
Higgs field. The unbroken $N=2$ SUSY corresponds to $\mu=0$ and $h=1$.

The properties of this broken $N=2$ theory have been considered previously
in [14 - 16]. The most detailed description has been given recently by
A. Gorsky, A. Vainshtein and A. Yung in [17], and we use widely the
results of this paper below. For our purposes, we will deal with the special
case of light quarks weakly coupled to the Higgs field:
\bq
m\ll\Lambda,\quad \sigma=h\left (\frac{\Lambda}{m}\right )^{3/2}\ll 1\,,
\eq
where $\Lambda$ is the scale parameter of the original fundamental theory
($\Lambda=1$ in what follows). Under the conditions of Eq.(10), one 
vacuum state decouples and there remain two physically equivalent vacuum
states. So, it will be sufficient to deal with one of them where the 
condensates of original fields take the values [17]:
\bq
\langle {\ov Q}Q\rangle \sim {\mu}\,{m^{-1/2}}\,;\quad 
\langle U\rangle=\langle \phi^2\rangle
\sim m^{1/2}\,;\quad\langle\lambda\lambda \rangle \sim \mu\,m^{1/2}\,,
\eq
while the condensate of the monopole field is
\bq
\langle {\ov M}M\rangle \sim \mu\, m^{1/4}\,.
\eq

Under the conditions of Eq.(10), the only freedom remained is the relative
value of $\mu$ and $m$, and the phase and physical content of this theory
depend essentially on this. Indeed:\\ {\bf a)} at 
sufficiently small $\mu\,;$ the quarks $Q\,(\ov Q)$ are "heavy" and
decouple, we are in the pure $N=2$ SYM - theory with $\Lambda_{eff}^{(1)}=m^
{1/4}\,\Lambda^{3/4}$, broken by the small $\mu\,U$ - term. The vacuum is 
the Seiberg-Witten vacuum, i.e. the dominant condensate is the Higgs one,
$\langle \phi^a \rangle$, leading to $SU(2)\ra U(1)$, with $W^{\pm}$ masses
$M_W\sim \Lambda_{eff}^{(1)}\sim m^{1/4}$. The light monopole field condenses
in the low energy $U(1)$ - theory, resulting in the confinement of electric 
charges. The lightest particles are the dual photon $\tilde \gamma$, its $N=2$ 
partner $A_D$ and the monopole composite $({\ov M}M)^{1/2}$, all with small 
masses $\sim \mu^{1/2} m^{1/8}$. 
We will call this phase the magnetic one.\\
{\bf b)} at sufficiently large $\mu\,;$ the Higgs field $\phi$ is heavy and 
decouples, we are in $N=1,\,N_F=1$ SQCD with $\Lambda_{eff}^{(2)}=\mu^{2/5}
\Lambda^{3/5}$ and with light quark composite $({\ov Q}Q)^{1/2}$- fields, 
with their masses $\sim m\ll \Lambda_{eff}^{(2)}$. Here, the large quark 
condensate $\langle Q_i\rangle \sim \langle {\ov Q}^j \rangle $ dominates, 
SU(2) is broken completely and there is confinement of magnetically charged 
excitations (monopoles), see the footnote 6.
The low energy effective Lagrangian is those of Affleck-Dine-Seiberg [11].
We will call this phase the electric one.

So, unlike the examples considered in previous sections, at the conditions 
given by Eq.(10) we have a good control here over the
phases of our theory in both limiting cases of small and large values
of $\mu$, and these phases are dual to each other and are dominated by
coherent condensates of mutually 
non-local monopole and quark fields. Our purpose now is to trace in 
more detail the transition between the magnetic and electric phases at some 
value $\mu \simeq \mu_o$, when going from small $\mu\ll \mu_o$ to large 
$\mu\gg \mu_o$ values of $\mu$. We expect that this transition proceeds 
through the formation of the mixed phase in some region $c_1\,\mu_o\leq \mu 
\leq c_2\,\mu_o$ (with $c_1 \leq c_2$, but parametrically both 
$c_1\sim c_2\sim O(1)$\,).~\footnote
{
Here and in other supersymmetric theories, the condensates of chiral 
superfields are frequently simplest smooth functions of chiral parameters 
{\it in the whole parameter space}, 
see for instance Eqs.(11),\,(12). This does not 
contradict to possibility for a system to be in qualitatively different 
(dual to each other) phases at different values of parameters, because these 
condensates are not order parameters for these phases. Rather, the masses of
the direct and dual photon (together with $W^{\pm}$ - masses) look more like 
the order parameters in the electric and magnetic phases respectively.
}  
\vspace{3mm}

In the magnetic phase region $0\leq\mu\lesssim\mu_o$ we will proceed in the 
same way as in the previous section, by retaining only the lightest fields
of the dual photon $\tilde \gamma$, $A_D$ and monopole $M$. All quark fields, 
in particular, are integrated out. Although, see Eq.(10), $h\ra 0$ and quarks 
do not interact directly with the Higgs fields, they interact with massive 
charged gluons and gluinos and will give corrections in powers of the
characteristic scale $\sim \langle \lambda\lambda \rangle /m^3 \sim (\mu/
m^{5/2})$. Further, being integrated out, the massive gluons and gluinos 
will transmit these corrections to the monopole $Z_M$-factor: $Z_M=Z_M(\mu
/m^{5/2})$. So, the quarks really decouple only at $\mu\ll m^{5/2}$ where 
$Z_M(\mu/m^{5/2}\ra 0)\ra 1$, while at $\mu > m^{5/2}$ the quarks influence 
the physics and $Z_M\neq 1$.

Similarly, in the electric phase region $\mu_o\lesssim \mu$, after 
integrating out the Higgs and gauge fields, the quark $Z_Q$-factor will 
obtain corrections in powers of the characteristic scale
$\sim \langle \lambda\lambda \rangle/\mu^3 \sim
(m^{1/2}/\mu^2)$, so that $Z_Q=Z_Q(m^{1/4}/\mu)$ and the heavy Higgs field
really decouples only at $\mu\gg m^{1/4}:\, Z_Q(m^{1/4}/\mu \ra 0)\ra 1$.
At $\mu_o\lesssim \mu < m^{1/4}$ the adjoint Higgs field still influences 
the physics, and $Z_Q\neq 1$.
\vspace{2mm}

It is not difficult to see that with the choice:
\bq
Z_{M}^{o}\equiv Z_M (\mu_o/m^{5/2} )\sim \frac{m^{1/4}}{\mu_o}\,\,,\quad 
Z_{Q}^{o}\equiv Z_Q (m^{1/4}/\mu_o ) \sim \frac{m}{\mu_o}\,\,,
\eq
all particle masses and all (properly normalized) condensates are matched
in the transition region $\mu\simeq \mu_o$, as it should be  (the gauge 
couplings are $\sim 1$ at $\mu\sim \mu_o$, see below):
\bq
M_H\sim M_M\sim M_Q\sim \mu_o\,;\quad M_{W^{\pm}}\sim M_{\gamma}\sim
M_{\tilde \gamma}\sim m^{1/4} > \mu_o\,,
\eq
\bq
\langle \phi^2 \rangle \sim \langle \sqrt{Z^o_Q}{\ov Q}\cdot
\sqrt{Z^o_Q}Q\rangle \sim \langle
\sqrt{Z_M^o}{\ov M}\cdot \sqrt{Z_M^o}M\rangle \sim m^{1/2}\,.
\eq
The nontrivial fact is that the number of matching conditions in Eqs.
(14) - (15) is larger than the choice of only two numbers in Eq.(13).

As for the value of $\mu_o$, it is determined
by matching at $\mu\sim\mu_o$ of two characteristic 
scales: those of $\Lambda^{(1)}_{eff}=m^{1/4}$ from the magnetic phase 
region $0 < \mu < \mu_o$, and those of $\Lambda^{(2)}_{eff}=\mu^{2/5}$ from 
the electric phase region $\mu >\mu_o\,$: 
\bq
m^{1/4}\sim \mu_o^{2/5} \quad \ra \quad \mu_o\sim m^{5/8}.
\eq

It is not difficult to see that this is equivalent to matching of the gauge 
couplings. Indeed, in the electric phase region the charged degrees 
of freedom decouple at the scale $\sim M_W=M_{\gamma}$, 
determined by the quark condensation, restructuring the spectrum 
and decoupling of the light neutral quark composite field $({\ov Q}Q)^{1/2}$. 
So, the inverse gauge coupling behaves as $\tau\sim\ln (M_W/
\Lambda^{(2)}_{eff})$, which is $\tau^o\sim \ln (m^{1/4}/\mu_{o}^{2/5})$ 
at $\mu\sim \mu_o$ (see Eq.(14)). 

In the magnetic phase region there are two characteristic scales: a) those 
connected with the Higgs condensation and decoupling of $W^{\pm}$ and gives 
$\tau\sim \ln (M_W/\Lambda^{(1)}_{eff})\sim 1$;\, and 
b) those connected with the monopole condensation and decoupling of the 
neutral monopole composite field $({\ov M}M)^{1/2}$. This last gives $\tau_D
\sim \ln (\Lambda^{(1)}_{eff}/M_{\tilde \gamma})$, which is also $\tau^o_D
\sim 1$ at $\mu\sim \mu_o$ (see Eq.(14)). Therefore, the matching of 
couplings at $\mu\sim \mu_o$ gives: 
$\tau^o\sim\tau^o_D\sim 1\sim\ln (m^{1/4}/\mu_{o}^{2/5})$.

So, $Z_M$ behaves like\,\, $Z_M = [1+(\mu/m^{5/2})]^{1/5}$ in the 
magnetic phase region $0 < \mu \lesssim m^{5/8}$,\,\, and the monopole 
mass (or, more exactly, the mass of the monopole composite field $({\ov M}M)
^{1/2}$) is: $M_M\sim \mu^{2/5}\,m^{3/8}$ at $m^{5/2} < \mu < m^{5/8}$, while
$Z_Q$ behaves like $Z_Q = [1+(m^{1/4}/\mu)]^{-1}$ in the electric phase 
region $m^{5/8}\lesssim \mu$. 

It is seen also from Eqs.(13) and (16) that there is a kind of duality 
relation at the transition point $\mu\sim \mu_o\,: Z_{M}^{o}\sim 1/Z_{Q}^{o}$.
Moreover, at small $\mu \ll m^{5/2}$ the 
confinement is very weak and the quark is nearly free, with mass $M_Q=m$ and
$Z_Q\ra 1$. So, $Z_{M}\sim 1/Z_{Q}$ holds also at small $\mu$ as well. 
Therefore, because the behaviour within the same phase region is smooth, the 
duality relation $Z_Q\sim 1/Z_M$ holds in the 
whole magnetic phase region $0< \mu < \mu_o$. 

Introducing the effective quark mass as: $M_Q^{(eff)}=Z_Q^{-1}\cdot m$, it is 
then: $M_Q^{(eff)}\sim Z_M\cdot m$,\, and $(M_M/M^{(eff)}_Q)\sim (\mu/
\mu_o)^{1/5} < 1\,$ at $m^{5/2}< \mu <\mu_o $.
This shows that quark is heavier than monopole in the whole magnetic phase 
region, and this is self consistent. 
\vspace{3mm}

As was pointed out above, the example considered in this section has an
advantage that we have a good control over the properties of the magnetic and 
electric phases at both sides, $\mu\ll m^{5/8}$ and $\mu\gg m^{5/8}$, of the 
transition region at $\mu\sim m^{5/8}$. As for the properties of the mixed 
phase in the transition region, they are similar to those described in
previous sections. In short:

1)\, At very small $\mu < m^{5/2}$ the condensates of the Higgs and
monopole fields dominate, with the coherent Higgs condensate responsible for
the $W^{\pm}$ masses $M_W$, and the coherent monopole condensate responsible 
for the dual photon mass $M_{\tilde \gamma}$.
The quarks are "heavy" and confined, and there are 
rare incoherent fluctuations of neutral quark-antiquark pairs. 

%In other words, the small chiral quark condensate $\langle{\ov Q}Q\rangle$ 
%is here not the "genuine" coherent condensate which gives masses 
%to gluons, but originates from the quantum loop effect: $\langle {\ov Q}Q
%\rangle \sim (\langle \lambda\lambda\rangle)/m$. 
%In its turn, the small gluino condensate is also not the genuine coherent
%condensate in this region, and can be thought as being induced by the 
%instanton(s): $\langle \lambda\lambda\rangle \sim \mu\,(\Lambda^{(1)}_
%{eff})^4/\langle \phi^2\rangle.$  

2)\, The density of these neutral quark pairs increases with increasing $\mu$.

3)\, These quark bags (or strings) percolate at $\mu=c_1\,m^{5/8}$. At $c_1\,
m^{5/8}\leq \mu\leq c_2\,m^{5/8}$, with $c_{1,2}\sim O(1)$, 
the system is in the mixed phase, where two 
infinite size connected coherent networks of electric and magnetic strings 
(or bags) coexist. In this region (parametrically): quark condensate $\sim$
Higgs condensate $\sim$ monopole condensate,\, quark mass $\sim$ Higgs mass 
$\sim$ monopole mass,\, photon mass $\sim$ dual photon mass, etc. (see Eqs.
(14)-(15)).

4) At $\mu=c_2\,m^{5/8}$ the quark condensate takes over and enforces
de-percolation of the connected coherent condensates of the monopole 
and Higgs fields. I.e., these last 
decay into separate independently fluctuating droplets, whose density and 
typical size decrease with increasing $\mu$, while the gluon masses
originate now from the coherent quark condensate.

5) At $\mu > m^{1/4}$ the Higgs field becomes too heavy and decouples 
completely, while the magnetically charged excitations are heavy
and confined into rare small size neutral pairs.   \\

\begin{center} {\bf 6.\,\,  Summary}  \end{center}

As has been argued on a number of examples above, the mixed phases exist with 
their properties qualitatively different from those of pure phases. 
And the appearance of mixed phases is not an exception but rather a typical 
phenomenon in various gauge theories, both supersymmetric and ordinary.\\

\begin{center}{\bf Conclusion}\end{center}

\begin{center} Dear Arkady, be healthy and happy\,! \end{center}
\vspace{5mm}

\begin{center}{\bf Acknowledgements}\end{center}

I am grateful to the Organizing Committee of "QCD-2002/Arkadyfest"
for a kind hospitality and support.

\end{document}